\documentclass[ twocolumn, aps,  superscriptaddress,showpacs, pre]{revtex4-1}
\usepackage{graphicx,amsfonts,amsmath,color,amsbsy,amssymb, subfigure}
\usepackage{epstopdf}
\usepackage{float}
\usepackage{xcolor}

\begin{document}
\title{The role of spontaneous curvature in the formation of cell membrane necks}
\author{G. Torres-Vargas}\email{gamaliel.tv@gmail.com} 
\affiliation{Posgrado en Ciencias Naturales e Ingenier\'ia.
Universidad Aut\'onoma Metropolitana Cuajimalpa.\\
Vasco de Quiroga 4871, 05348 Cd. de  M\'exico, MEXICO}


\author{F. Monroy }\email{monroy@quim.ucm.es}
\affiliation{Departamento de Qu\'imica F\'isica\\ 
Universidad Complutense de Madrid\\
Av. Complutense s/n, 28040, Madrid, SPAIN}
\affiliation{Institute for Biomedical Research Hospital Doce de Octubre (imas12)\\
Av. Andaluc\'ia s/n 28041, Madrid, SPAIN}


\author{J.A. Santiago }\email{jsantiago@cua.uam.mx}
\affiliation{Departamento de Matem\'aticas Aplicadas y Sistemas\\ 
Universidad Aut\'onoma Metropolitana Cuajimalpa\\
Vasco de Quiroga 4871, 05348 Cd. de  M\'exico, MEXICO}
\affiliation{Institute for Biomedical Research Hospital Doce de Octubre (imas12)\\
Av. Andaluc\'ia s/n 28041, Madrid, SPAIN}

\vspace{10pt}


\begin{abstract}
The mechanical effects of  membrane compositional inhomogeneities  are analyzed in a 
process  analogous of neck formation in cellular membranes.
We cast on  the Canham-Helfrich model  of fluid membranes with both the  spontaneous curvature 
and the surface tension  being non-homogeneous functions along the cell membrane. The inhomogeneous distribution is determined by the equilibrium mechanical equations, and, in order to  establish the  role played by the inhomogeneity, we focus on the catenoid, a surface of zero mean curvature, which can be described in terms of the 
catenary curve parameterized by arc length. We show that  analytic solutions exist for the  spontaneous curvature, as well  
as for both,  the surface tension and the radial elastic force. An analytic expression for the constrictive  force at the neck,
is obtained. From the energetic analysis, it is found that, if we fix the value of the constrictive force at the neck, 
the set of solutions lies on two branches separated by an energetic barrier. This barrier corresponds to the energy of the maximum catenoid.
If instead we fix the axial force,  the solution has access to catenoid  of any size.

\smallskip
\end{abstract}


\maketitle

\section{Introduction}
Biological membranes are ubiquitous in nature as an essential component of Life. For all living cells, and many biological compartments, a fluid mosaic membrane constitutes the flexible envelope indispensable for adopting a functional form at selective exchange of mass and energy with the environment \cite{SINGER}. Such a functional fitness of biological membranes should stem as a direct causality between their mesoscopic structure -arisen from the fundamental properties of mosaicity and fluidity, and the way they function -at interaction with the surrounding. The fluid mosaic  is constituted  by tens of lipid species providing structural and functional  support to numerous proteins executing very specific biochemical tasks. Being fluid and mosaic upon functional necessity, biological membranes are multicomponent systems evolutionally optimized with a chemical heterogeneity entailed by the many molecular interactions between their molecular ingredients. Indeed, biological membranes are nowadays recognized to be {\it more mosaic than fluid} \cite{ENGELMAN},  
with a natural heterogeneity (mosaicity) sustaining differentiated regions of form and function,
\cite{BEN-MUN, MAC}. This membrane mosaicity is highly variable across different cellular emplacements, even varying at dynamic dependence of the 
mechanical and biomechanical status along the cell cycle. Membrane inhomogeneity is indeed increasingly recognized as a key regulating factor in cell shape remodelling processes.

In this work, we invoke spontaneous  curvature as regulated by specific proteins\cite{MAC}, and by  lipids too\cite{KZ}, 
to  actively impart the local stress needed  for executing the mechanical program of membrane remodeling. Membrane tension is also known to vary in response to local changes in membrane area \cite{LIPOWSKY2014}, a fact probably related to sensing mechanisms exploited by cell physiology at mechanical connection between membrane and cytoskeleton forces. Despite these  biological complexities, the physical paradigm still depicts the flexible membrane as a homogeneous sheet able to adapt its curved form to an isotropic class of cellular hydrostatics \cite{BOAL}. Membrane curvature has been historically considered as a continuous geometry  abstracted by a homogeneous elasticity field accounting for bending and tensional stresses exerted by the cell \cite{BOAL}. 

The Canham–Helfrich (CH) theory is the cornerstone for cell  mechanics as defined for homogeneous membranes \cite{LIPOWSKY-NATURE, SEIFERT-ADV}. Building upon the mean curvature $(K)$  as  systemic invariants for local shapes, the canonical CH-model describes membrane energetics as a harmonic curvature-elasticity field subjected to homogeneous material constraints. The CH-field is completely isotropic as far considers the fluidity characteristic to endow  homogenous material properties (taking exactly the same value at every site in the membrane). Assuming constant values for the membrane properties i.e. 
rigidity ($\kappa$), surface tension ($\sigma$) and spontaneous curvature ($K_0$), 
this classical CH-model describes the changes in membrane energy without compositional  heterogeneities \cite{SEIFERT-ADV}
\begin{equation}
{\cal H}=  \frac{\kappa}{2} \int dA (K-K_0)^2 + k_g \int dA H +  \sigma \int dA\  \label{CAHEL}, 
\end{equation}
where $K$ and $H$ represent, respectively,  mean and Gaussian curvatures characterized by constant moduli ($\kappa$ and $k_g$).
Despite  the achievements of this homogeneous theory describing cellular shapes with simplified vesicle models \cite{LIPOWSKY-NATURE, SEIFERT-ADV},  the mechanical role of biological mosaicity has not been investigated so far. 
Under the extended hypothesis of a relevant material heterogeneity at mechanical crosstalk with membrane geometry, we address here the inhomogeneous version of the CH-theory. Compositional heterogeneity (mosaicity) is enabled at mechanical equilibrium through of local connections with membrane geometry, which leave off the systemic non locality endowed by the fluidity property in homogeneous CH-elasticity. 

We consider inhomogeneous material properties with a variable spatial dependence determined by local values of $K_0$ and $\sigma$ curvatures, which are ultimately imposed by the precise molecular composition at every membrane emplacement. Under such inhomogeneity hypothesis,   we are especially 
interested on  how less mechanical work is required for cell-like heterogeneous vesicles to undergo shape transformations with respect to the homogeneous case.

\section{Inhomogeneous membrane  equilibrium}\label{BST}

{\it Surface geometry}. A generic surface in $\mathbb{R}^3$,  with cartesian coordinates ${\bf  x}= (x^1, x^2, x^3)$, can be  parametrized by the embedding 
function  ${\bf x}= {\bf X}(\xi^a)$, where $\xi^a$ are local coordinates on the surface $ (a=1, 2)$. The infinitesimal 3D-Euclidean distance $ds^2=d{\bf x}\cdot d{\bf x}$, induces the corresponding arc length distance on the surface $ds^2= g_{ab}d\xi^a d\xi^b$;  here $g_{ab}={\bf e}_a \cdot {\bf e}_b$ is the 
induced metric, and  ${\bf e}_a=  \partial_a {\bf X}$ are two local tangent vector fields. Correspondingly, the induced metric defines a covariant derivative  on the surface denoted by $\nabla_a$.
The unit normal to the surface is ${\bf n}\,=  (\varepsilon^{ab}/2) {\bf  e}_a \times{\bf e}_b $, where 
$\varepsilon^{ab}= \epsilon^{ab} / \sqrt{g} $
with $ \epsilon^{ab}$ being the Levi-Civita alternating simbol and $g= \det \left(g_{ab}\right)$.
The Gauss equation stablishes $ \nabla_a {\bf e}_b= -K_{ab}{\bf n}$, which describes the change of  the tangent vector fields along the surface. 
The components of the extrinsic curvature are defined as $K_{ab}= -\nabla_a {\bf e}_b\cdot {\bf n} $,  which are related with the 
intrinsic Gaussian curvature  ${\cal R}_ G$ through of 
the Gauss-Codazzi equation, $K_{a}^cK_{cb}=KK_{cb}-g_{ab}{\cal R}_G$, and its contraction $K^{ab}K_{ab}= K^2-2{\cal R}_G$ \cite{willmore}.   
The Codazzi-Mainardi equation $\nabla_a K^{ab}=\nabla^b K$,  will be also useful~\cite{willmore}.

\smallskip
{\it Elastic energy}. As defined by the  Canham-Helfrich functional describing mean-field curvature elasticity \cite{canham,helfrich}; for the
inhomogeneous case, this is:
\begin{equation}
{\cal H}=  \frac{\kappa}{2} \int dA (K-K_0)^2 + \int dA \, \sigma, \label{CAHEL}
\end{equation}
where $\kappa$ is the constant (globally averaged) value of bending rigidity, and  $dA$  the area element. 
Both, the spontaneous curvature $K_0$, and the surface tension ($\sigma $) are considered inhomogeneous functions i.e., they are variable quantities that depend
on the local stress. As a strong constraint for membrane heterogeneity,  the local surface tension $\sigma$ has been introduced as  a coordinate-dependent  Lagrange multiplier, which locally fixes the surface area  of each membrane element. No Gaussian term is considered as conserved along the considered
shape transformations (without topological change: Gauss-Bonnet theorem \cite{willmore}).

Let's notice  the inhomogeneous state-variables $\sigma (\xi^a)$, and $K_0 (\xi^a)$  as accounting 
for the local values of lateral tension, and spontaneous curvature, respectively ($\xi^a$ denotes surface coordinates). 
As a reference state, we consider a homogeneous sphere of radius $R$, which represents the ground state as far as a constant  spontaneous curvature  coincides with its natural curvature, i.e. $K_0=2 / R$. 
In the event that such spherical vesicle  is deformed without topological charge, inhomogeneities appear in spontaneous curvature  and
lateral tension as a mechanical response to maintain  equilibrium.

\smallskip
{\it Stress tensor.}  The stress distribution along the membrane is the information encoded in the stress tensor, given by  \cite{jem, auxiliary, fournier},
\begin{equation}
{\bf f}^a= f^{ab} {\bf e}_b + f^a {\bf n},  \label{STR}
\end{equation} 
where the tangential and normal components are:
\begin{eqnarray}
f^{ab}&=& \kappa (K-K_0)\left[ K^{ab} - \frac{1}{2}(K-K_0) g^{ab} \right]  - g^{ab}\sigma, \nonumber\\
f^a &=&-\kappa \nabla^a (K-K_0). \label{PRJJ}
\end{eqnarray}
For closed membranes (spherical topology) the  hydrostatic pressure term $-PV$,  is added to the energy in Eq. \eqref{CAHEL}; $P=P_{in}-P_{out}$ is the 
pressure jump, and $V$ the enclosed volume.
At mechanical equilibrium,  the surface divergence of the stress tensor is thus written as
\begin{equation}
\nabla_a {\bf f}^a = P\, {\bf n}.\label{EQRR}
\end{equation}
Therefore, substituting Eq. \eqref{STR} into Eq. \eqref{EQRR}, the equilibrium conditions  in terms
of the stresses in Eqs. \eqref{PRJJ} as
\begin{eqnarray}
\nabla_a f^a -K_{ab} f^{ab}&=&P,\nonumber\\
\nabla_a f^{ab}+ f^a K_a{}^b&=& 0.\label{ELL}
\end{eqnarray}
These equilibrium equations  define a generalized framework for calculating  constitutional relationships for the inhomogeneous membrane (mosaic membrane), which determine local shape in terms of local elasticity (as given by variable elastic parameters). 
\smallskip

{\it Inhomogeneous CH-membrane: local shape equations.}  Specifically, Eq. \eqref{PRJJ} provides analytic expressions for the local stresses compatible with the inhomogeneous CH-membrane, as considered at global mechanical equilibrium. By substituting with the particular expressions of Eq. \eqref{PRJJ}, in the generalized equilibrium equations of Eq. \eqref{ELL}, one immediately gets the inhomogeneous connections between local elasticity and curvatures.

First condition in \eqref{ELL} accounts for  mechanical stability along  the normal direction:
\begin{eqnarray}
&-&\kappa \nabla^2 (K-K_0) - \frac{\kappa}{2} (K-K_0) \Big[K(K+ K_0) - 4 {\cal R}_G \Big]\nonumber\\
&+& \sigma K=P, \label{NORMAL-EQ}
\end{eqnarray}
which describes the connection between  local shape and Laplace pressure at any point in the membrane;
In the homogeneous case (for constant $\sigma$ and $K_0$), Eq. \eqref{NORMAL-EQ} reduces to the 
well-known Helfrich's shape equation  \cite{oyang, oyang1}.

The second condition in Eq. \eqref{ELL}  accounts for the lateral equilibrium equation, or the tensile stress in the tangent plane; this is:
\begin{equation}
\partial_a \sigma =\kappa (K-K_0)\partial_a K_0.\label{TANGT}
\end{equation}
This result is particularly interesting as connects in a linear relationship the change in surface tension with the change in spontaneous curvature as it happens with respect to the local value of mean curvature. 
Being novel in the analytical form here presented, the concept is quite intuitive in biological terms, and extremely useful for interpreting 
the impact of curvature makers in membrane tension.

{\it Boundary conditions.} For open membranes the boundary conditions are given by \cite{SANTIAGO}
\begin{eqnarray}
\kappa (K-K_0) &=&  -k_g K_T, \label{BU} \\
\frac{\kappa}{2} (K-K_0)^2  + \sigma &=&  -k_g {\cal R}_G, \label{BUUB} \\
\kappa \nabla_l K &=& k_g K'_\tau.\label{BUBB}
\end{eqnarray}
Eq. \eqref{BU} implies that,  on the boundary, the torque is proportional to the projection of 
the extrinsic curvature, $K_{ab}$,  on the tangent vector; being the constant of proportionality,  
the ratio between the Gaussian stiffness and the bending modulus, $ - k_g/  \kappa$. 
The second boundary condition, Eq. \eqref{BUUB},  identify the  energy density 
with the Gaussian curvature, ${\cal R}_G$.

{\it Axial summetry}.  For membranes with axial symmetry,  the Euler-Lagrange equations can be  obtained from the Helfrich functional in cilindrical coordinates. Following 
\cite{SEIFERT-1}, we have ${\cal H}= 2\pi  \int ds \, L $,
where
\begin{eqnarray}
L &=&\frac{\kappa \rho}{2} \left( \Psi' + \frac{\sin\Psi}{\rho} - K_0  \right)^2 + \sigma \\
&+& {\gamma}(\rho' -\cos\Psi) + {\eta} ( z' + \sin\Psi),
\end{eqnarray}
where $\gamma$ and $\eta$, are Lagrange multipliers, that enforces the relationship with the tangential angle, $\Psi$, and $'$, denotes
derivative respect to the arc length, $s$, which parametrizes the generating curve.

\section{The catenoid}
\begin{figure}[H]  
\centering 
\includegraphics[scale=0.35]{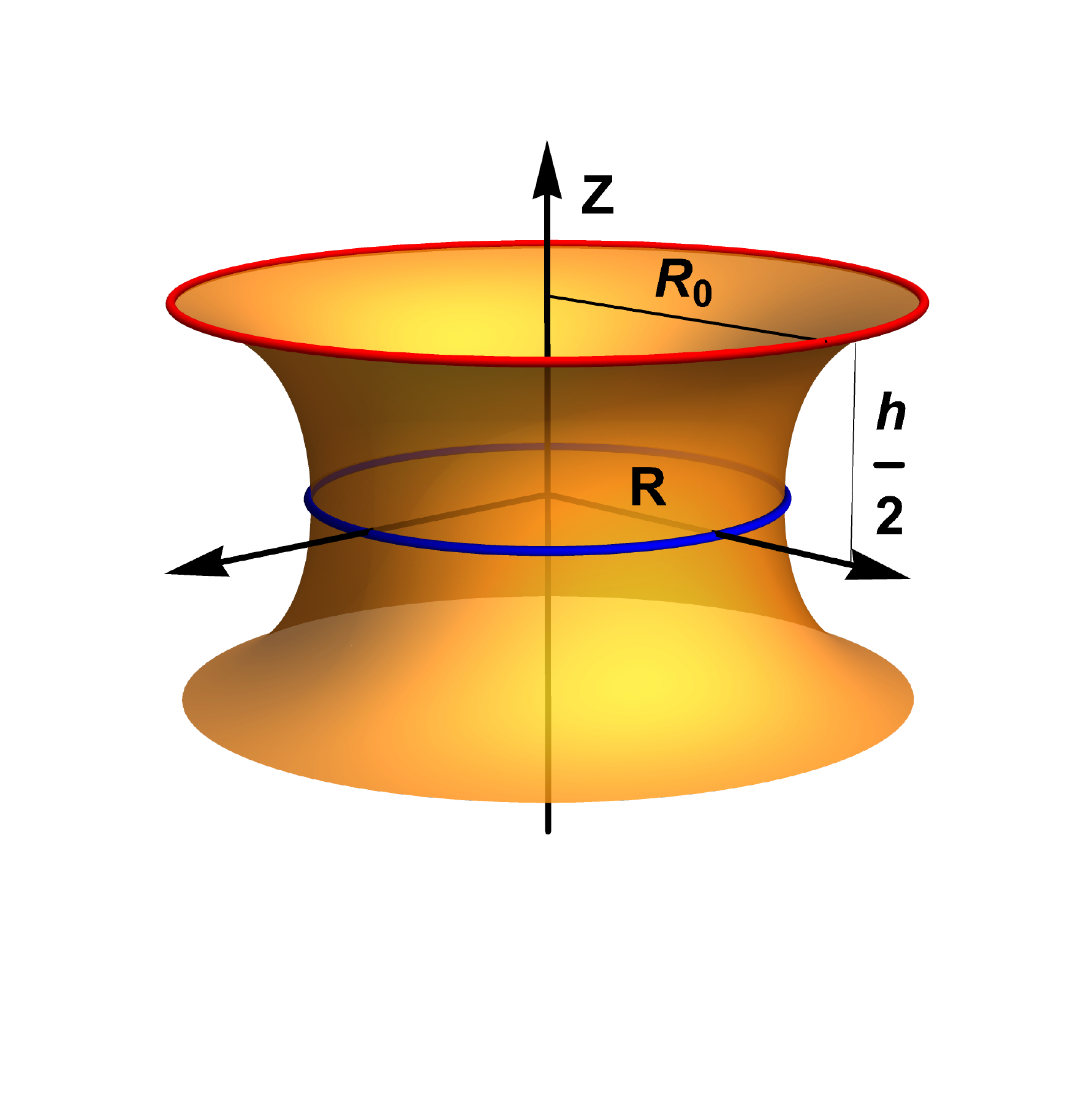}
\vspace{-15mm}
\caption{ Defining the dimensionless variables, $\bar \rho \equiv \rho/ R$, $\bar z \equiv z/ R$,  the equation of the catenoid with neck radio  $R$, is given by $\bar \rho=  \cosh \bar z$,  so that $\sinh \bar z= -\bar l $,
where  $\bar l \equiv l/R - L/(2R)$.  On the upper border (the initial point), $\bar l=\bar l_b= -\bar L/2$, on the neck, $\bar l=0$, and in the lower border, $\bar l= -\bar l_b$. Therefore, the catenoid is parametrized in terms of $\bar l$, through the functions
$ \bar\rho =  \sqrt{1+\bar l^2},$ $\bar z = - {\rm ArcSinh}\,  \bar l $. On the upper border we have $\bar R_0^2 (\bar l_b) =1 +\bar l_b^2$.
The relationship with the tangential angle of the generating curve:
$\sin\Psi = 1/  \sqrt{1+ {\bar l}^2} $,  and
$\cos\Psi = \bar l /\sqrt{1+ {\bar l}^2}$. The derivative respect to $\bar l$, is given by, 
$\bar\Psi' = -1/ (1+\bar l^2)$.
}
\label{CAT-1}
\vspace{-2mm}
\end{figure}

This  is a minimal surface with curvature, $K=0$ (we parametrize this surface as shown in Fig. \ref{CAT-1}).
In terms of the dimensionless variables,  
$\bar \rho\equiv \rho/R$, $\bar \gamma \equiv \gamma R/\kappa$, $\bar \eta \equiv \eta R/\kappa$, 
$\bar\sigma \equiv \sigma R^2/ \kappa$, the mechanical balance, along the axial and radial directions are  given respectively by 
\begin{eqnarray}
\bar\rho \cos\Psi \bar K_0' + \bar\rho \sin\Psi \, \Psi' K_0 + {\cal C}_0 \bar\rho \sin\Psi= - \bar\eta,\label{FIR}\\
\bar\rho\sin\Psi \bar K_0'- \bar\rho \cos\Psi\, \Psi'\bar K_0  - {\cal C}_0 \bar \rho \cos\Psi = - \bar\gamma.
\end{eqnarray}
Thus,   $-\bar\gamma$ 
plays the role of an external force that balances the radial  elastic force, whereas the negative of  $\bar\eta$, balances the axial elastic force.
In this case, Eq. \eqref{TANGT}, can be integrated directly, so that
the dimensionless surface tension, results to be
\begin{equation}
 \bar\sigma (\bar l) = - \bar K_0^2(\bar l) /2 + {\cal C}_0,\label{SIS}
\end{equation} 
where ${\cal C}_0$ is a constant of integration. This constant can be determined 
by the boundary condition, Eq. \eqref{BUUB}, as ${\cal C}_0= - (k_g /\kappa) \bar {\cal R}_g(\bar l_b)$, where $\bar l_b$, denotes arc length parameter ${\bar l}$,  at the boundary, i.e.  
$\bar l_b=-L/(2R)$ (see Fig. \ref{CAT-1}).

Using  the parametric equations  for the catenoid (see the caption in Fig. \eqref{CAT-1}), 
the axial balance,  Eq. \eqref{FIR}, is  written as 
\begin{equation}
\bar l \bar K_0' (\bar l)-\frac{\bar K_0(\bar l)} {1+\bar l^2 }  + {\cal C}_0= -\bar\eta.\label{PER}
\end{equation}
Solutions to this equation,   give the spontaneous curvature along the catenoid as
\begin{equation}
\bar K_0 (\bar l) =  (\bar \eta +{\cal C}_0) - (\bar\eta + {\cal C}_0)  \,  {\cal G}(\bar l)  +  \frac{ {\cal C}_1\,  \bar l  }{  \sqrt{1+ {\bar l}^2}},   \label{SOP}  
\end{equation}
where ${\cal C}_1$ is a constant of integration, and  we have defined the function
\begin{equation}
{\cal G}(\bar l)\equiv \frac{\bar l} {\sqrt{1+ {\bar l}^2}   }  {\rm ArcTanh}\left( \frac{ \bar l}{  \sqrt{1+ {\bar l}^2} }\right).
\end{equation}
The constant ${\cal C}_1$, is obtained by evaluating  Eq. \eqref{SOP} on the upper  border, where $\bar l=\bar l_b$. In terms of the scaled radio 
$\bar R_0\equiv R_0/R$, we get
\begin{equation}
 \bar{\cal C}_1 =  \frac{  1  }{ \bar R_0 \sqrt{\bar R_0^2  -1} } +  \frac{ \bar R_0 }{ \sqrt{ \bar R_0^2 -1} }  \label{CC1}
\left(\mu -  \frac{1 }{\bar R_0^4(\bar l_b)} \right)  [ {\cal G}(\bar R_0)-1 ],
\end{equation}
where we have defined,  $\bar{\cal C}_1\equiv {\cal C}_1 \kappa/ k_g$,  and   $\mu\equiv \bar \eta   \kappa/ k_g$.
In Eq. \eqref{CC1},  the only root of the function in brackets, 
$\bar R_0^*\simeq 1.81$, so that, for this value of the upper border,  the function $\bar{\cal C}_1 \simeq  0.366$, does not  depend  on the $\mu$-parameter. 
If,  in Eq. \eqref{CC1}, we fix the value of $\mu$ and substitute it in Eq. \eqref{SOP}, we will have the distribution of the spontaneous curvature on a catenoid, with 
upper boundary ${\bar R}_0$, and the end point to be determined.

In order to identify the location $\bar l_B$, of the  lower border, with  radial parameter $\bar R_B$,  we impose the boundary condition 
Eq. \eqref{BUUB}, on the spontaneous curvature, Eq. \eqref{SOP}, in such away that,  $\bar K_0 (\bar l_B) = (k_g/\kappa)\bar R_B^{-2}$.
It implies that
\begin{eqnarray}
\frac{ 1}{ \bar R_B^2}  &=& \left(  \mu - \frac{1}{\bar R_0^4} \right) \left(\bar R_B^2-  2\bar A_B  \frac { \sqrt{\bar R_B^2 -1}  } { \bar R_B  }    \right)\nonumber\\
&+& \bar{\cal C}_1  \frac { \sqrt{\bar R_B^2 -1}  } { \bar R_B  }, \label{TU}
\end{eqnarray}
where, $\bar A_B\equiv A_B/(2\pi R^2) $, is the area of south hemisphere of the catenoid, scaled with the neck radio, $R$. For the  lower hemisphere of  the catenoid,  we obtain
\begin{equation}
2{\bar A_B}= \bar R_B \sqrt{\bar R_B^2-1 }  + {\rm ArcTanh} \left(  \frac{ \sqrt{\bar R_B^2-1} }{\bar R_B }  \right), \label{ER}
\end{equation}
and a similar formula  for the upper hemisphere, with $\bar A_0$,  and border radio, $\bar R_0$.

After substituting $\bar{\cal C}_1$ into Eq. \eqref{TU}, and simplifying, is not difficult to prove that the only solution to Eq. \eqref{TU}, is given by  $\bar R_B=\bar R_0$, 
so that,  $\bar l_B=-\bar l_b$.
Therefore, the boundary conditions on the spontaneous curvature,  $\bar K_0(\bar l)$,  force the catenoid to be a  symmetric shape; this does not mean that the 
function $\bar K_0(\bar l)$, be symmetric. Indeed, the last term in Eq. \eqref{SOP},  tells us that this is not the case, except if ${\cal C}_1=0$.

Evaluation of Eq. \eqref{SOP}, on $\bar l=0$, gives the spontaneous curvature  on the neck:
\begin{equation}
\bar K_0^{neck}= \frac{k_g}{\kappa}\left( \mu -\frac{1}{\bar R_0^4(\bar l_b)}\right).
\end{equation}

Regarding the surface tension, Eq. \eqref{SIS}, it is obtained as
\begin{equation}
\bar\sigma({\bar l})= -\frac{\bar K_0^2(\bar l)}{2} - \frac{k_g}{\kappa} \frac{1} {{\bar R} _0^4(\bar l_b) }. \label{ENE}
\end{equation}
On the border we get
\begin{equation}
\bar\sigma (\bar l_b) = -\frac{k_g}{\kappa} \frac{1}{2\bar R_0^4(\bar l_b)} \left( 2+\frac{k_g}{\kappa}  \right). 
\end{equation}
The radial force, $-\, \bar \gamma$,  is found to be
\begin{equation}
-\bar\gamma(\bar l) = \bar K_0'(\bar l)  +\frac{\bar l}{ 1+ \bar l^2}\bar K_0(\bar l) -\bar l\,  {\cal C}_0.\label{CONST}
\end{equation}
The constrictive force on the neck, is obtained by evaluating  Eq. \eqref{CONST}, at $\bar l=0$, and therefore $-\bar\gamma (0) =\bar K_0' (0)= {\cal C}_1$, so that
the constrictive force is given by
\begin{eqnarray}
-\bar\gamma (0) \frac{\kappa}{k_g} &=&  \frac{1}{ \bar R_0 \sqrt{\bar R_0^2 -1}  }   \nonumber\\
&+& \frac{ \bar R_0 }{ \sqrt{\bar R_0^2-1  }} \left(\mu -  \frac{1 }{\bar R_0^4 } \right)  [ {\cal G}(\bar R_0) -1 ]. 
\end{eqnarray}
It is worth writing this equation, in terms of the neck radio,  scaled with $R_0$, that is, $r_N\equiv R/R_0$. Thus, the constrictive scaled force, $\tilde F_C\equiv -\gamma (0)R_0/\kappa$, (modulo $\kappa/k_g$),  
is given by
\begin{eqnarray}
\tilde F_C\frac{\kappa}{k_g}	&=& \frac{r_N}{ \sqrt{1-r_N^2}} \nonumber\\
&+& \frac{1}{  \sqrt{1-r_N^2  }} \left(\tilde \eta \frac{\kappa}{k_g}  - r_N^3 \right)  [ {\cal G}(r_N) -1 ],
\end{eqnarray}
where we have defined,  $\tilde\eta \equiv \eta R_0 /\kappa$, and ${\cal G}(r_N)\equiv \sqrt{1-r_N^2} \rm{ArcTanh} [ 1/ (  \sqrt{1-r_N^2}) ]  $.
Therefore,  the constrictive force on the neck,  does depend on the neck radio, $ r_N$, as well as  the scaled axial force, $ - \tilde \eta$.

Of course, for the catenoid with $\bar R_0^*\sim 1.81$, the constriction force does not depend on the $\mu$-parameter.
To see the meaning of this catenoid, which is the  intersection point of all the curves with fixed $\mu$, let us write the area but scaled with the boundary radio, $R_0$, i.e., 
$\tilde {A}\equiv A/( 2\pi R_0^2)$, and thus $ \tilde A = \bar A / \bar R_0^2$, where $\bar A$, is given by the r.h.s. of Eq. \eqref{ER}. It can be seen that the area  
$\tilde A$, reaches its maximum value at the point,  $\bar R_0^*\simeq 1.81$. Nevertheless, as we shall see, this is not the catenoid of minimum energy.

\begin{figure}[!htbp]  
\centering 
\includegraphics[scale=0.18]{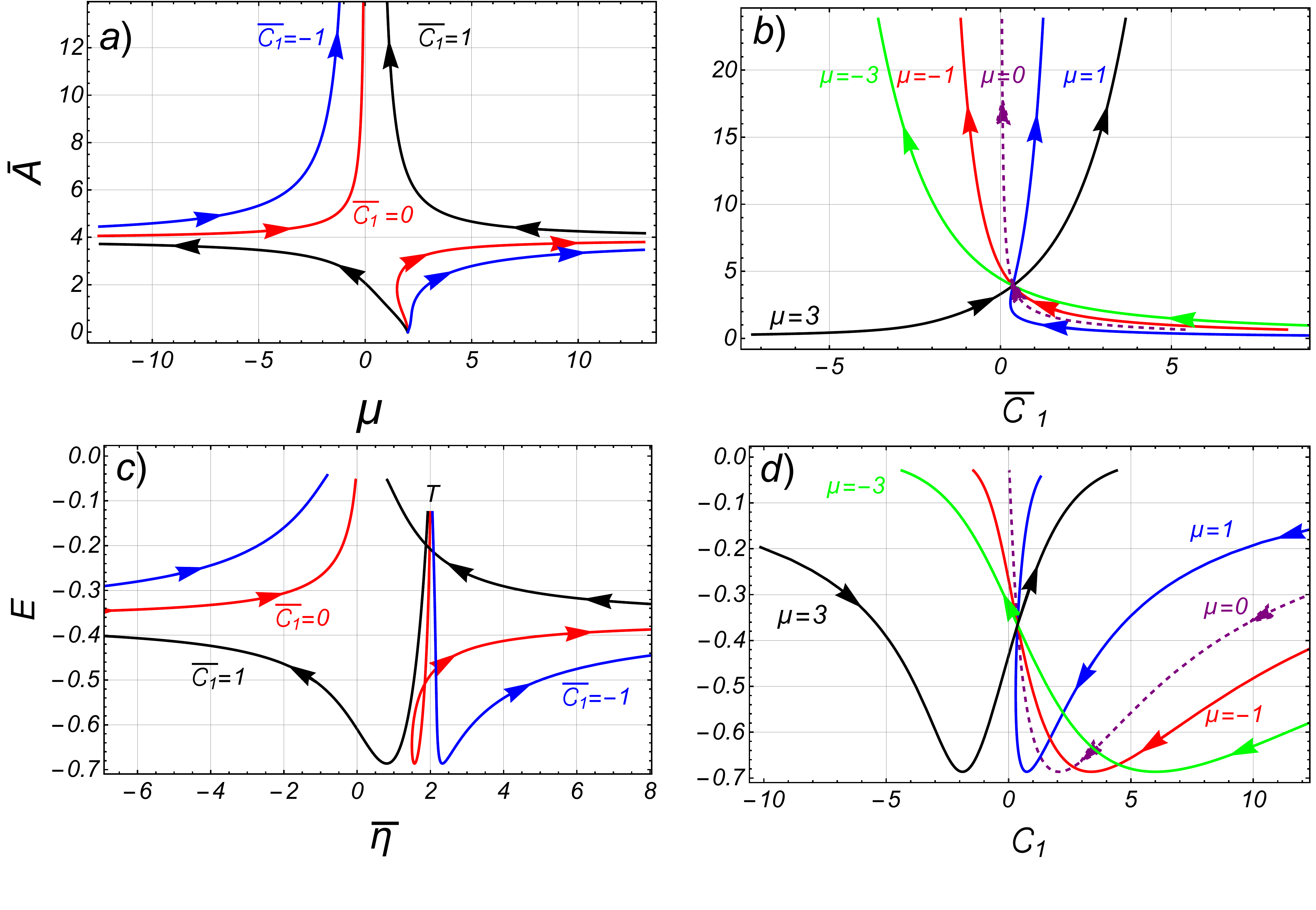}
\vspace{-5mm}
\caption{ $a)$ The parametric-plot, $(\bar A (\bar R_0),  \mu (\bar{\cal C}_1 , \bar R_0 ) )$. 
Points on the curves represent catenoids such that,  $\bar{\cal C}_1\equiv (\kappa/k_g) {\cal C}_1$, 
is a constant. The arrows indicate increasing values of $\bar R_0$ (decreasing values of the neck radio, $r_N$).
The diagram is separated into two regions by the horizontal line, $\bar A^* \simeq 3.93$, where, $\bar R_0^*\simeq 1.81$. 
The catenoid cannot move, classically,  from configurations of one region to another, by these trajectories. 
$b)$ Curves  of constant  $\mu$, obtained from the parametric-plot, $(\bar A (\bar R_0), \bar{\cal C}_1 (\mu , \bar R_0 ) )$. 
The intersection point represents the maximum catenoid, reached at $\bar R_0^*$, and $\bar A^*$.
$c)$ The energy as a function of  $\bar \eta$, and  the corresponding trajectories in the panel $a)$: if the catenoid starts from a certain point $T$, 
with higher energy, any small change in $\bar\eta$, moves the catenoid to its point of minimum energy, $E_{min}\simeq -0.68$, with  ${\bar R_0}\simeq 1.176$.
If we continue to change the $\bar\eta$-parameter, the energy of the catenoid increases but not beyond a certain point, where
there is an energetic  barrier, $E^*\simeq -0.366$, which corresponds to the maximum catenoid energy; 
but if initially the catenoid is far from the point $T$, then, by modifying the value of $\bar\eta$, the catenoid evolves to 
configurations with energy  higher than $E^*$. 
$d)$ The energy,  $E$, as a function of ${\cal C}_1$ and the corresponding  trajectories  in the panel $b)$; 
the intersection point represents the  maximum catenoid. 
}
\label{MAP-1}
\vspace{-2mm}
\end{figure}

According to Eq. \eqref{ENE}, the energy of the catenoid is given by 
\begin{equation}
E =- \frac{k_g}{\kappa} \frac{\bar A} {{\bar R} _0^4},
\end{equation}
where $\bar A$, is the r.h.s of Eq. \eqref{ER}. Then, the functions $\bar A(\bar R_0)$, and $\mu (\bar C_1, R_0 )$, are parametrized
by $\bar R_0$. The resulting plots are showed in Fig. \ref{MAP-1}a),b).
Similarly, the parametric plots of  the energy $E(\bar R_0)$ vs. ${\cal C}_1$ and $\mu$   are depicted in Fig. \ref{MAP-1}c),d).

In  Fig.  \ref{MAP-1}$a)$,  solutions with fixed $\bar {\cal C}_1$ (the constraint force on the neck),  
are showed to be  on two branches. These
are separated by the horizontal  asymptote, $\bar A^*\simeq 3.93$, that appears as a singularity   in the function, $\mu= \mu (\bar {\cal C}_1, {\bar R}_0)$, 
such that ${\cal G}(\bar R_0^*)=1$, i.e. $\bar R_0^*\simeq 1.81$. Precisely, $\bar R_0^*$ and $\bar A^*$, get the maximum values 
that the catenoid could reach. 
According to this result, the catenoid can not transit, classically,  from one branch to the other, by trajectories
of constant $\bar{\cal C}_1$: If $\bar A <  \bar A^*$, and  we increase  the magnitude of  $\mu$, then $\bar A$ slowly
increases, approaching  asymptotically to $\bar A^*$. On the other hand, if  $\bar A \gtrsim \bar A^* $, and we decrease the magnitude of  $\mu$, the
parameter $\bar A$ increases,  lowering the energy, as showed in the panel $c)$, where the corresponding  energy of the 
trajectories in Fig. \ref{MAP-1}$a)$, is shown.

If we fix the $\mu$-parameter, the results are shown in Fig. \ref{MAP-1}$b)$, and the corresponding values of the energy,  in the panel $d)$.
In both figures, the intersection point corresponds to the maximal size of the catenoid,  
Nevertheless, as we observe in the panel $d)$, this it is not the point of minimal energy.

\begin{figure}[!htbp]  
\centering 
\includegraphics[scale=0.18]{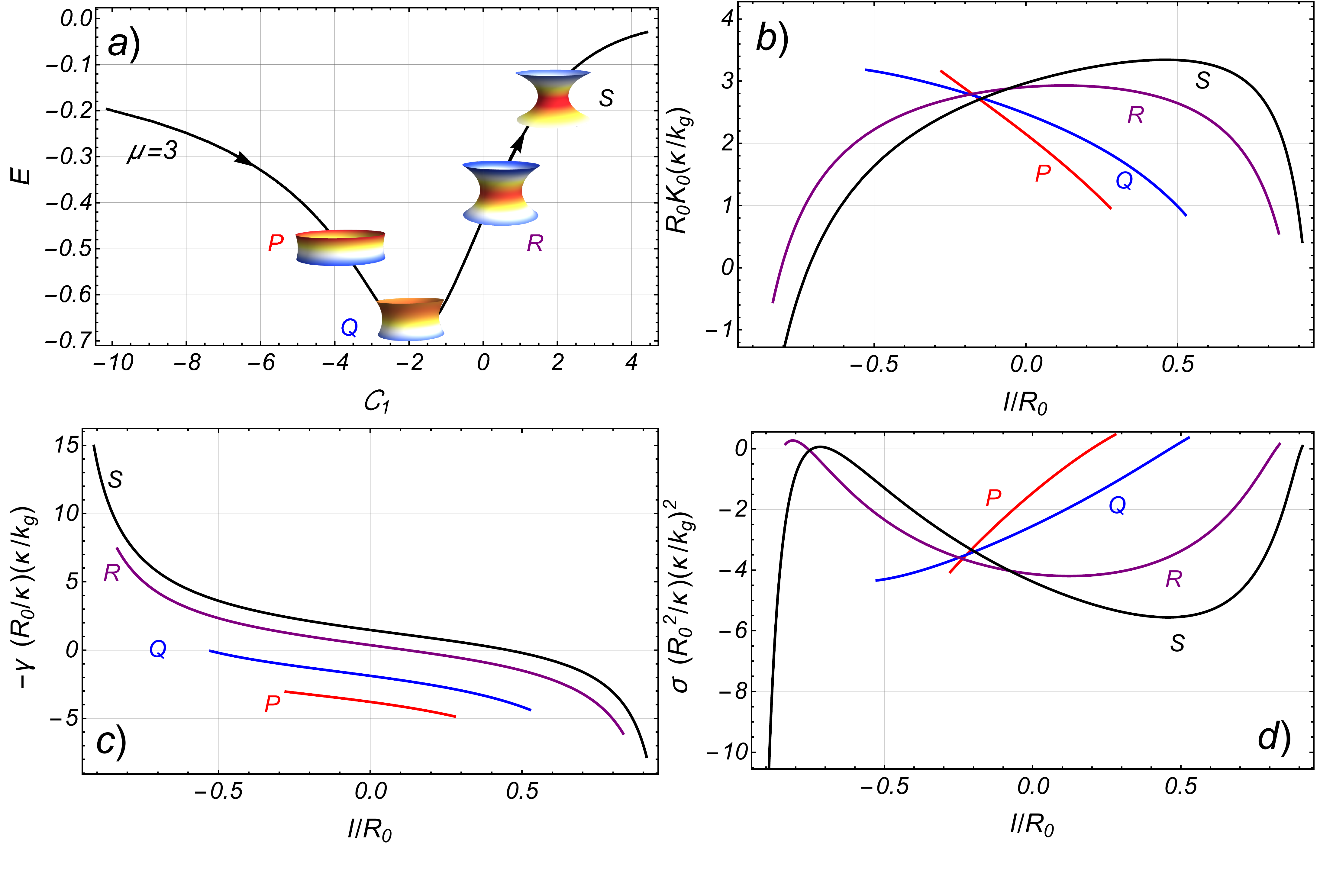}
\vspace{-5mm}
\caption{ a) Some specific solutions in the case $\mu\equiv \bar\eta (\kappa/k_g)=3$. The arrow means increasing values of the parameter, $\bar R_0 \equiv R_0/R$ 
(decreasing values of the neck radio, $r_N$).
The catenoid at $P$, with reduced area,  $A/(2\pi R_0^2)\simeq 0.54 $, reduced height,  $h/R_0 \simeq 0.54$, and neck radio, $r_n\equiv  R/R_0=0.96.$ 
At  $Q$, corresponds the minimal energy with:  $A/(2\pi R_0^2)\simeq 1.06 $, $h/R_0 \simeq 0.99$, and $r_n\simeq 0.85$.
At the point $R$, corresponds the maximum size, $A/(2\pi R_0^2)\simeq 1.19 $, $h/R_0 \simeq 1.32$  and $r_n\simeq 0.55$. 
At $S$, we have: $A/(2\pi R_0^2)\simeq 1.17 $, $h/R_0 \simeq 1.26$, and $r_n\simeq 0.41$.
b) Distribution of the spontaneous curvature (note the factor, $\kappa/k_g$), as a function of $l/R_0$. On the  R-catenoid,  the function is almost symmetric.
c) Distribution of the radial force, as a function of $l/R_0$. Note the negative 
values,  of the constrictive force on the neck, at $P$ and $Q$, and the positive values at $R$ and $S$.
d) Distribution of the surface tension as a function of $l/R_0$.
}
\label{MAR}
\vspace{-2mm}
\end{figure}

Some specific values have been depicted in Fig. \ref{MAR}, on the path with $\mu=3$ (see Fig. \ref{MAR}a). The Q-catenoid with the minimal energy, and the R-catenoid with
the maximal values of relative area and height (see Fig. \ref{MAR}b). On the  catenoids P and Q, the neck is under a negative ${\cal C}_1$, while at R and S, it is under a positive one (see Fig. \ref{MAR}c). On the neck of the P-catenoid, we have,  $- \tilde\gamma (0) \kappa/k_g\simeq -3.79 $, thus the dimensionless  force on the neck is,  $-\tilde \gamma (0)_P\simeq 2.65$, 
where  we have taken, $k_g/\kappa=-0.7$. Similarly, the constriction force on the Q-catenoid is,  $-\tilde \gamma (0)_Q\simeq 1.31$. On the maximal,  R-catenoid
$-\tilde \gamma (0)_R\simeq -0.25$.  Finally, on the S-catenoid
$-\tilde \gamma (0)_S\simeq -1.03$. Thus, on this curve, where the axial force remains fixed, we will have higher energy catenoids as the neck of the catenoid closes or opens, respect to 
the neck-radio of the minimal energy, $r_N\simeq 0.85$.

Note that, possible oscillations between catenoids of different neck radii should be observed, to achieve this, in the first case, (see Fig. \ref{MAP-1}c), 
the total energy must be less than $E^*$, oscillations in the axial force must be implemented. In the second case,    (see Fig. \ref{MAP-1}d), it can be 
implemented by oscillating the radial force and in both cases, it is possible to realize it by modifying the distribution in the spontaneous curvature.

\section{Summary and discussion}

A problem of great interest is the theoretical or experimental description of the forces involved in the formation of invaginated corrugated forms and in endocytosis or exocytosis processes in cell membranes. For example, in the process of cell division, among the multiple factors, passive and active, that participate, it is important to distinguish the basic elements that play a relevant role.
In this sense, to describe the elastic forces involved, the theoretical model proposed by Helfrich in the 1970s has been very useful. One of the phenomenological elements, introduced to take into account possible asymmetries on the membrane, is what is called spontaneous curvature. These asymmetries induce a certain preferred curvature, spontaneously, by the membrane. It is natural then, to suppose that these asymmetries are not homogeneous, although a lot of progress has been achieved under the hypothesis that the spontaneous curvature is homogeneous \cite{beltran-heredia, PABLO}, and very little has been said in the inhomogeneous case \cite{chabanon}.
 The hypothesis that we adopt in this work is that both the spontaneous curvature and the surface tension are inhomogeneous functions. To isolate the effect of spontaneous curvature we have focused on the catenoid, a surface of zero mean curvature. Another advantage of doing so is that, by the cylindrical symmetry, the analysis is centered on meridians of the catenoid.
Indeed, we have found that the {\it spontaneous constriction force}, acting on the neck of the catenoid,  is the derivative of the spontaneous curvature at that point, which in turn has an analytical expression, solution to the axial balance equation. \\
Our analysis reveals that, if we fix the constricting force at the neck of the membrane, starting with a large radius, then, if we decrease the neck radius $r_N$, modifying the axial force, the possible solutions approach asymptotically to the energy of the maximum catenoid, so that we do not have access to catenoids with very thin necks. This defines a first branch of solutions, in which a very interesting phenomenon also occurs with spontaneous curvature: It goes from having positive to negative concavity, like a phase transition, a fact that is directly reflected in the radial force.  The second branch of the solutions allows us to achieve very thin neck catenoids, for this, we would have to start from a catenoid with a neck radios smaller than  $r_{N-max}$. 
In a few words, there is an energetic barrier that do not allow us to move from one branch of solutions to the other.
If instead we fix the value of the axial force, we can go towards very thin neck catenoids, modifying the value of the constriction force on the neck.

Possible oscillations between catenoids of different neck radii should be observed. To achieve this, in the first case, Fig.\ref{MAP-1}c), oscillations in the axial 
force should be implemented. In the second case, Fig.\ref{MAP-1}d), it can be achieved by oscillating the radial force and in both cases, 
it is possible to implement it, by modifying the distribution in the spontaneous curvature. In fact, happily we have seen that recently an oscillatory phenomenon of this nature has been reported for the case of the neck formed for a budding transition, in which modifications of the spontaneous curvature have been made \cite{PETRA}, 
moreover, an alternative theoretical analysis to the one presented here has been made to explain it \cite{PETRA-1}.

The framework here developed  will allow us to advance our understanding of  {\it spontaneous forces} in other 
problems, including, for example, the estomatocyte-discocyte transition.

\smallskip

\smallskip

\section*{Acknowledgements}
JAS whishes to thank Gil Barrientos and Jairo Bustamante for helpful discussion,  GTV would like to thank CONACyT 
for support through a scholar fellowship (Grant No 381047).


\end{document}